\documentclass[twocolumn]{article}
\usepackage{booktabs}							
\usepackage{nicefrac}							
\usepackage{textcomp}							
\usepackage{upgreek}							
\usepackage{mathrsfs}							
\usepackage{bm}										
\usepackage{slashed}							
\usepackage[utf8]{inputenc}       
\usepackage[german,english]{babel}        
\usepackage{amssymb}							
\usepackage{amsmath}							
\usepackage{dsfont}               
\usepackage{braket}								
\usepackage{graphicx}							
\usepackage{color}								
\usepackage{transparent}					
\usepackage{float}								
\usepackage[titletoc]{appendix}		
\usepackage{multicol}							
\usepackage{authblk}

\newcommand{\im}{\mathrm{i}}
\newcommand{\D}{\mathscr{D}}
\newcommand{\halb}{\tfrac{1}{2}}
\newcommand{\ihalbe}{\tfrac{\mathrm{i}}{2}}
\begin{document}
\twocolumn[

\title{On topologically massive gravity with extended supersymmetry}
\date{}
\author{Frederik Lauf${}^{\dagger}$ and  
Ivo Sachs${}^{\dagger*}$}
\affil{${}^{\dagger}$Arnold Sommerfeld Center for Theoretical Physics, Ludwig-Maximilians-Universit\"at\\
Theresienstra\ss{}e 37, D-80333 M\"unchen, Germany\\
${}^{*}$Department of Applied Mathematics and Theoretical Physics, University of Cambridge\\ Cambridge, CB3 OWA, UK}
\begin{@twocolumnfalse}
	\maketitle
	\begin{abstract}
	We describe the construction of $2+1$-dimensional toplogically massive adS gravity with ${\mathcal{N}}$-extended supersymmetry in superspace by means of introducing a compensating hypermultiplet for the super-Weyl invariance. For $\mathcal{N}\geq 3$ the scalar multiplet must be on shell and the potential for the scalar compensator is completely determined by the geometry. As a consequence the resulting massive theory has no free parameter for $\mathcal{N}\geq 4$. For $\mathcal{N}= 4$ we show that this leads to topologically massive gravity at the chiral point and construct the corresponding off-shell component action.
	\end{abstract}
	\vspace{2\baselineskip}
\end{@twocolumnfalse}
]
\section{Topologically massive gravity}
The simplest theory of gravity in three dimensions with a massive, propagating degree of freedom is topologically massive gravity, defined by the action \cite{C1}
\begin{equation*}
S=S_{\mathrm{EH}}+\frac{1}{\mu}S_{\mathrm{top}}\,,
\end{equation*}
where 
\begin{equation*}
S_{\mathrm{EH}}=-\frac{1}{\kappa^{2}}\int{\mathrm{d}^3x~e\left(R+2\ell^{-2}\right)}
\end{equation*}
is the  Einstein-Hilbert action and
\begin{equation*}
S_{\mathrm{top}}=\frac{1}{\kappa^{2}}\int{\left(\omega\wedge R-\tfrac{2}{3}\omega\wedge\omega\wedge\omega\right)}
\end{equation*}
describes a gravitational Chern-Simons term. The latter action is invariant under Weyl rescalings of the metric, $g_{\mu\nu}\to e^{2\sigma(x)}g_{\mu\nu}$ and therefore, on its own, has no propagating degree of freedom in $2+1$ dimensions in spite its higher derivative kinetic term. The combination $\lambda=\mu\kappa^2$ is the Chern-Simons coupling. The negative cosmological constant is related to the anti-de Sitter radius $\ell$ through $\mathit{\Lambda}=-\ell^{-2}$. In view of its embedding in supergravity we shall use the customary realization of the Einstein-Hilbert term with the help of a conformal compensator $\phi$, coupling to the Riemann scalar together with a potential
\begin{equation*}
S_{\mathrm{cc}}=\int{\mathrm{d}^3x~e\left(-\halb\partial^a\phi\partial_a\phi-\halb\xi R\phi^2+c\phi^6\right)}.
\end{equation*}
For conformal coupling $\xi=\tfrac{1}{8}$. The compensator $\phi$ can be set to any non-vanishing  constant by a suitable Weyl transformation. Fixing this constant to a given value breaks Weyl invariance spontaneously and furthermore fixes the value of Newton's constant $\kappa^2$. The resulting theory, which propagates one massive degree of freedom, is characterized by the dimensionless quantity $\mu\ell$ and has been shown to violate unitarity or positivity for $\mu\ell\neq 1$ \cite{Strominger}. More precisely, since the conformal compensator leads to the "wrong" sign of Newtons constant the perturbative modes have positive energy while black holes generically have negative energy (see also \cite{Carlip:2008jk} and \cite{Grumiller:2008qz} for a detailed discussion of stability). However, as we shall see below, in $\mathcal{N}\geq 4$ superconformal gravity the $\mu\ell$ is automatically fixed when using a scalar hypermultiplet as the conformal compensator.
\section{$\mathcal{N}$-extended scalar multiplets}
Let us start with the algebra of supercovariant derivatives
\begin{equation*}
\{{\D}_\alpha^I,{\D}_\beta^J\}=\im\updelta^{IJ}(\gamma^a)_{\alpha\beta}{\D}_a+\tfrac{1}{2}R^{IJ,KL}_{\alpha\beta}\mathcal{N}_{KL}\,,
\end{equation*}
acting on a scalar hypermultiplet $\underline{Q}$ (e.g. \cite{Bergshoeff:2010ui}), which will play the role of the compensator superfield and which is taken to transform in the fundamental representation of $\mathrm{spin}(\mathcal{N})$ with generators $\mathcal{N}_{KL}$. Acting with a supercovariant derivative on it gives a spinor superfield 
\begin{equation*}
{\D}_\alpha^I\underline{Q}=\underline{\mathit{\Lambda}}_\alpha^I\,.
\end{equation*}
The superderivative of the latter can be parametrized as \cite{Howe}
\begin{equation*}
{\D}_\alpha^I\underline{\mathit{\Lambda}}_\beta^J=(\gamma^a)_{\alpha\beta}\underline{H}_a^{IJ}+\upvarepsilon_{\alpha\beta}\underline{H}^{IJ}.
\end{equation*}
For this parametrization to be consistent with the algebra on $\underline{Q}$ via the relation
\begin{equation*}
\{{\D}_\alpha^I,{\D}_\beta^J\}\underline{Q}={\D}_\alpha^I\underline{\mathit{\Lambda}}_\beta^J+{\D}_\beta^J\underline{\mathit{\Lambda}}_\alpha^I
\end{equation*}
one must have
\begin{equation*}
\underline{H}_{(\alpha\beta)}^{(IJ)}=\ihalbe\updelta^{IJ}{\D}_{(\alpha\beta)}\underline{Q}+\tfrac{1}{4}R^{(IJ),KL}_{(\alpha\beta)}{\cal{N}}_{KL}\underline{Q}
\end{equation*}
\begin{equation*}
\upvarepsilon_{\alpha\beta}\underline{H}^{[IJ]}=\tfrac{1}{4}R^{[IJ],KL}_{[\alpha\beta]}{\cal{N}}_{KL}\underline{Q}\,,
\end{equation*}
while $\underline{H}_a^{[IJ]}$ and $\underline{H}^{(IJ)}$ are arbitrary. The algebra on the spinor superfield can be similarly constructed from the parametrization above. Focusing on the term giving rise to the torsion
\begin{align*}
\{{\D}_\alpha^I,{\D}_\beta^J\}\underline{\mathit{\Lambda}}_\gamma^K
=&\im(\gamma^a)_{\alpha\beta}\updelta^{K(I}{\D}_a\underline{\mathit{\Lambda}}_\gamma^{J)}
\end{align*}
one can see that the correct $\mathrm{SO}(\mathcal{N})$ index structure is obtained if the form of  $\underline{\mathit{\Lambda}}_\gamma^J$ is
\begin{equation*}
{\D}_\alpha^I\underline{Q}=\mathit{\Sigma^{I}}\underline{\mathit{\Lambda}}_\alpha
\end{equation*}
where $\mathit{\Sigma^{I}}$ are the $\mathrm{SO}({\mathcal{N}})$ spin matrices. This constraint is also intuitive because $\underline{\mathit{\Lambda}}_\alpha$ should be related to superpartners of $\underline{Q}$. The above equations become
\begin{equation*}\label{dl}
{\D}_\alpha^I\underline{\mathit{\Lambda}}_\beta=(\gamma^a)_{\alpha\beta}\underline{H}_a^I+\upvarepsilon_{\alpha\beta}\underline{H}^I
\end{equation*}
with
\begin{equation*}
\mathit{\Sigma^{(J}}\underline{H}_{(\alpha\beta)}^{I)}=\ihalbe\updelta^{IJ}{\D}_{(\alpha\beta)}\underline{Q}+\tfrac{1}{4}R^{(IJ),KL}_{(\alpha\beta)}{\cal{N}}_{KL}\underline{Q}
\end{equation*}
\begin{equation*}
\upvarepsilon_{\alpha\beta}\mathit{\Sigma^{[J}}\underline{H}^{I]}=\tfrac{1}{4}R^{[IJ],KL}_{[\alpha\beta]}{\cal{N}}_{KL}\underline{Q}\,.
\end{equation*}
Thus the SUSY variation of $\underline{\mathit{\Lambda}}_\alpha$ is fixed except for $\mathcal{N}=1$  and $\mathcal{N}=2$ (the latter is described by an $\mathcal{N}=1$ algebra of complex supercharges).   For $\mathcal{N}\geq 3$, hypermultiplets can then only exist on shell due to the consistency of the derivative of $\underline{\mathit{\Lambda}}_\alpha$ with the algebra on $\underline{Q}$. Furthermore, the resulting equations of motion for $\underline{\mathit{\Lambda}}_\alpha$ and $\underline{Q}$ are fully determined in terms of the geometry encoded in the algebra of the supercovariant derivatives. For $\mathcal{N}=1$ the field ${H}$ is arbitrary and can be used as an auxiliary field with appropriate supersymmetry transformation in order to close the  algebra on $\underline{\mathit{\Lambda}}_\alpha$ off shell.\footnote{Off-shell actions for $\mathcal{N}=1,2$ topologically massive supergravity can be found in \cite{Deser:1982sw} and \cite{Kuzenko:2013uya} respectively. $\mathcal{N}=3,4$ topologically massive supergravity with vector multiplet compensators can be found in \cite{Kuzenko:2014jra}. Off-shell supergravity-matter couplings have been developed in \cite{Kuzenko:2011xg} for $\mathcal{N}\leq 4$. See also \cite{Bergshoeff:2010ui} for a discussion of linearised massive SUGRA.}
\section{Superconformal gravity}
In supergravity the R-symmetry is local and must be gauged, giving rise to an $\mathrm{SO}(\mathcal{N})$ field strength $F_{ab}^{IJ}$. For $\mathcal{N}\geq 4$ the latter is contained in a totally antisymmetric $\mathrm{SO}(\mathcal{N})$ superfield $M^{IJKL}$ \cite{howe0} that contains in addition the Cotton tensor -- the three dimensional analogue of the Weyl tensor (e.g. \cite{C1}). More precisely \cite{Kuzenko:2011xg,Howe}, 
\begin{align*}
R^{IJ,KL}_{\alpha\beta}=&\im\upvarepsilon_{\alpha\beta}\left(M^{IJKL}+4\delta^{[K[I}K^{J]L]}\right)\nonumber\\
&-\im(\gamma^a)_{\alpha\beta}\left(4\delta^{[K(I}L_a^{J)L]}-\delta^{IJ}L_a^{KL}\right)\,.
\end{align*}
The fields $K$ and $L$ belong to a compensating Weyl multiplet since their leading components can be gauged away by super-Weyl transformations \cite{howe0,Kuzenko:2011xg}. 

Upon coupling the compensator hypermultiplet to conformal supergravity (i.e. the $\cal{N}$-fold  supersymmetrization of $S_{\mathrm{top}}$ \cite{Butter:2013rba}), the on-shell field strength is determined by the conformal compensator current through the equation of motion for the $\mathrm{SO}({\cal{N}})$ gauge field
\begin{equation*}
\varepsilon^{abc}F_{bc}^{IJ}\propto \lambda\left(\phi^\dagger\mathit{\Sigma^{[IJ]}}\partial^a\phi-\partial^a\phi^\dagger\mathit{\Sigma^{[IJ]}}\phi\right).
\end{equation*}
Here $\phi$ is the first component of the hypermultiplet $Q$. 
Thus $M^{IJKL}$ (and therefore the geometry modulo Weyl transformations) is determined by the matter superfield, that is 
\begin{equation*}
M^{IJKL}\propto\lambda~Q^\dagger\mathit{\Sigma^{[IJKL]}}Q.
\end{equation*}
The precise coefficients are fixed by the projection on the third component which is related to the field strength via the algebra of supercovariant derivatives. 
 
On the other hand we have seen in the previous section that the equation of motion and therefore the potential for $\phi$ will be determined by the geometry via the supersymmetry algebra, more precisely by $M^{IJKL}$ which therefore is related to the cosmological constant $\ell^{-2}$ generated by the scalar potential. This is the mechanism through which $\mu\ell$ is fixed for ${\mathcal{N}}\geq 4$. Using various methods including the one described here it was shown in \cite{Chu,Howe} that the superconformal gauging of ABJM and BLG models leads to a discrete set of possible values for $\mu\ell$ for $\mathcal{N}=6$ and  $\mathcal{N}=8$ respectively. 

We will now show that $\mu\ell=1$ for $\mathcal{N}=4$ with a single scalar compensator \cite{Lauf}. For $\mathcal{N}=4$ we have $M^{IJKL}=\varepsilon^{IJKL}M$. Since we are looking for a maximally symmetric solution we will set $\underline{H}_a^{I}=0$. Furthermore, there must be a super-Weyl frame in which the modulus of the compensator $\phi$ is constant which is compatible with the SUSY algebra only if 
\begin{equation*}\label{hm}
\underline{H}^{I}=-\tfrac{1}{2}M\mathit{\bar{\Sigma}}^I \underline{Q}+K^{IJ}\mathit{\bar\Sigma}_{J}\underline{Q}=0.
\end{equation*}
Thus $K^{IJ}=\tfrac{1}{2}M\delta^{IJ}$.\footnote{In \cite{Kuzenko:2012bc} it was shown that the square of the symmetric matrix $K^{IJ}$ must be proportional to the identity for an anti-de Sitter superspace. Our solution corresponds to $(4,0)$ adS geometry in this classification.} Here the compensator is a left-handed $\mathrm{SU}(2)$ spinor. Had we chosen a right-handed compensator instead, $M$ would be replaced by $-M$. Note, however, that  in contrast to the linearised theory of \cite{Bergshoeff:2010ui}, the closure of the SUSY algebra allows only one non-vanishing, constant compensator on-shell. The cosmological constant can then be read off from the dimension two commutator
\begin{equation*}
[{\D}_a,{\D}_b]=4K^2{\cal{M}}_{ab}
\end{equation*}
which is indifferent to the spinor handedness. Finally we determine $M$ in terms of $\underline{Q}$. For this we compare the equation of motion for the $\mathrm{SO}(4)$ field strength 
\begin{equation*}
F_{ab}^{IJ}=-\frac{\lambda}{16}\varepsilon_{abc}\left[(\D^c\phi)^\dagger\mathit{\Sigma}^{IJ}\phi-\phi^\dagger\mathit{\Sigma}^{IJ}\D^c\phi\right]+\cdots\,,
\end{equation*}
where the periods stand for terms involving fermions, with the projection on $F_{ab}^{IJ}$ in $M$ \cite{Butter:2013rba}, 
\begin{equation*}
F_{ab}^{IJ}=\frac{i}{4}\varepsilon_{abc}(\gamma^c)^{\beta\gamma}\D^{[I}_\beta \D^{J]}_\gamma M|.
\end{equation*}
From this we infer that
\begin{equation*}
M=-\frac{\lambda}{16}|Q|^2
\end{equation*}
which in turn fixes the cosmological constant through $M=2K$, that is 
$$\ell^{-2}=w^2 =\left(\tfrac{1}{16}\right)^2\mu^2\kappa^4\phi^4$$
where $w=M|$ is the lowest component of $M$ and $\kappa^2=16\phi^{-2}$. Thus $\mu\ell=1$. At this point, we should comment on possible generalizations by gauging a flavor group for the compensator. Closure of the SUSY algebra admits two possibilities, namely $\mathrm{U}(n)\times\mathrm{U}(m)$ and $\mathrm{O}(m)\times\mathrm{Sp}(2n)$ with the compensator in the bifundamental representation \cite{Gaiotto:2008sd,Kuzenko:2015lfa}. However, for this gauged compensator the only covariantly constant solution is $|Q|^2=0$ which leaves us with pure superconformal gravity and vanishing cosmological constant \cite{Lauf}. Thus $\mu\ell=1$ is the unique solution. 

To conclude we present the off-shell action for $\mathcal{N}=4$ supersymmetric topologically massive gravity
\begin{align*}
S=\mkern-4mu\int\mkern-2mu\mathrm{d}^3x~\tfrac{e}{\lambda}\Big[&\tfrac{\upvarepsilon^{abc}}{2}(R_{ab}\omega_c-\tfrac{2}{3}\omega_{a}\omega_{b}\omega_{c})-\tfrac{\im}{8}f^c\gamma_d\gamma_cf^d\\
&-\tfrac{\upvarepsilon^{abc}}{2}(F_{ab}B_c+\tfrac{2}{3}B_aB_bB_c)\\
&-8\im w_I^\alpha w^I_\alpha-2wy\\
&-4\im\chi_{aI}\gamma^aw^Iw-\ihalbe\upvarepsilon^{abc}\chi_{a,I}\gamma_b\chi_c^{\phantom{c}I}w^2\Big]\\
-\tfrac{e}{\kappa^2}\Big[& R-\ihalbe\chi_{a}f^a+\tfrac{1}{2}B^a_{ij}B_a^{ij}+2y+2w^2\Big].
\end{align*}
The first four lines belong to the topological action taken from \cite{Butter:2013rba},  $w_\alpha^I$ and $y$ are auxiliary fields in the superconformal gravity multiplet $M$ \cite{howe0,Butter:2013rba}, $f^a=2\upvarepsilon^{abc}\D_b\chi_c$, and $B_{ij}=-\halb(\mathit{\Sigma}^{IJ})_{ij}B_{IJ}$ is the self-dual part of the gauge fields.

\section*{Acknowledgments}
We would like to thank Sergei Kuzenko, Bengt Nilsson, Joseph Novak and Martin Rocek for helpful discussions. I.S. would like to thank DAMTP at Cambridge University  for hospitality during the final stages of this work. This work  was supported by the DFG Transregional Collaborative Research Centre TRR 33 and the DFG cluster of excellence "Origin and Structure of the Universe".

\end{document}